# Zero-Poisson's-Ratio Elastomeric Substrates for Distortion-Free Stretchable Displays

JOSEPH NGUYEN,[†] KIM TRAN,[†] TRISTAN TJUSSARDI,[†] TIGER LIANG, ERIC YI, ASAD NAUMAN, AND ABDOULAYE NDAO,[*]

*Department of Electrical and Computer Engineering, University of California, San Diego, La Jolla, CA 92093, USA*

[†]*These authors contributed equally.*

**a1ndao@ucsd.edu*

**Abstract:** Stretchable displays are critical for emerging wearable electronics, soft sensors, and next-generation AR/VR interfaces. Although recent advances have enabled foldable, twistable, and rollable displays, intrinsically stretchable substrates often exhibit significant lateral contraction under tensile strain due to their high Poisson's ratio, leading to unintended wrapping, distortion, and shrinkage. Here, we report a transparent heterogeneous-modulus elastomeric substrate designed to achieve near-zero Poisson's ratio while maintaining mechanical softness and optical transparency. The substrate consists of line-patterned hard polydimethylsiloxane (PDMS) embedded within a soft PDMS matrix, producing spatially heterogeneous strain distribution during stretching. In this architecture, the soft PDMS functions as a strain-absorbing medium, while the embedded hard PDMS patterns suppress lateral deformation perpendicular to the applied strain. As a result, the structure significantly dampens transverse contraction and realizes a near-zero effective Poisson's ratio. To demonstrate the utility of this platform for stretchable optoelectronics, LED arrays were integrated onto the heterogeneous substrate. The devices exhibit minimal vertical and lateral distortion during tensile deformation, enabling mechanically stable operation of stretchable light-emitting displays. This heterogeneous modulus strategy provides a simple, scalable approach to mechanically robust stretchable display platforms.

## 1. Introduction

The emergence of next-generation electronic systems, encompassing skin-conformable health monitors, soft robotics, electronic textiles, and augmented and virtual reality (AR/VR) interfaces, has created an urgent demand for display platforms capable of accommodating large and repeated mechanical deformations while preserving optical and electrical functionality [1-5]. Mechanically deformable displays, including flexible, foldable, rollable, and stretchable variants, have received considerable attention due to their broad applicability across diverse electronic systems, with stretchable displays representing the most advanced form factor [6-8]. Unlike flexible or foldable displays, which deform along a single bending axis, stretchable displays must withstand multidimensional tensile strain without compromising image quality, pixel integrity, or mechanical reliability, requirements that place fundamentally new demands on the design of their constituent substrates [9-12].

The progression from rigid to flexible, and subsequently from foldable to rollable displays, has been enabled by transitioning from glass to thin plastic substrates and then to elastomeric platforms [13,14]. Flexible displays on thin plastic substrates have been commercialized, and foldable and rollable displays have been introduced, allowing users to adjust screen size on demand. However, achieving true in-plane stretchability introduces a mechanical challenge qualitatively distinct from bending: when an elastomeric substrate is stretched uniaxially, it contracts significantly in the perpendicular direction because of its Poisson's ratio [15]. Most stretchable displays are made from highly elastic elastomeric materials, but these materials exhibit a positive Poisson's ratio, leading to unavoidable image distortion when the display is stretched. For typical soft elastomers such as polydimethylsiloxane (PDMS), the Poisson's ratio approaches 0.5—the theoretical limit for an incompressible material—meaning that stretching the substrate by a given strain induces nearly half as much transverse shrinkage [16,17]. This lateral contraction leads to pixel displacement, aspect ratio distortion, warping of displayed content, and mechanical stress concentrations at device junctions, all of which compromise display performance and durability.

Several strategies have been proposed to overcome the Poisson’s ratio problem in stretchable substrates. One prominent approach exploits auxetic mechanical metamaterials, geometrically patterned structures that exhibit near-zero and even negative Poisson's ratios through their unique structural deformation mechanisms [18-24]. Auxetic metamaterials are a unique class of materials with a wide array of functionalities; however, their inherent porosity poses challenges for practical applications and filling the inherent perforations while preserving their auxetic nature is difficult because it demands the seamless integration of components with highly distinct mechanical characteristics [25-27]. Realizing these structures typically requires multi-material processing, precision patterning of rigid frameworks within soft matrices, and careful interfacial engineering, which adds significant fabrication complexity and raises concerns about delamination under cyclic loading [28-33]. Another approach leverages composite architectures in which stiff reinforcing elements—such as aligned fibers, ribbons, or metamaterial frames—are embedded within a compliant elastomeric matrix [34-37]. Continuously aligned transparent ribbon arrays incorporated within a stretchable matrix can confer mechanical anisotropy, suppressing lateral displacement when devices are stretched perpendicular to the ribbon alignment and enabling near-zero Poisson's ratio behavior. While effective, such composite strategies involve chemical-level heterogeneity—introducing dissimilar polymers, glass fabrics, or rigid inclusions—which complicates processing, can generate interfacial stress mismatches, and may reduce optical transparency through scattering at material interfaces.

A key insight motivating the present work is that Poisson’s ratio of a composite substrate can, in principle, be controlled not by introducing chemically distinct materials, but by engineering spatial heterogeneity in mechanical modulus within a chemically homogeneous elastomeric system. PDMS is an ideal platform for this strategy, as its elastic modulus is highly tunable over a wide range by simply varying the base-to-crosslinker ratio during preparation, with the elastic modulus varying linearly with crosslinker concentration, spanning from sub-MPa to several MPa depending on the mixing ratio [38-41]. Crucially, both soft and hard PDMS formulations share the same siloxane chemistry, ensuring strong covalent bonding at their interface upon sequential curing and eliminating the delamination risk inherent in multi-material composites. Furthermore, because both phases have similar refractive indices and optical properties, the resulting substrate maintains high optical transparency—a critical requirement for display applications.

Building on this principle, we introduce a heterogeneous-modulus elastomeric substrate composed entirely of PDMS, in which periodically spaced line patterns of high-modulus PDMS are embedded within a continuous matrix of low-modulus PDMS. Under applied tensile strain, the soft PDMS matrix acts as a strain-absorbing medium, accommodating most of the elongation, while the hard PDMS line patterns, aligned perpendicular to the stretching direction, resist lateral contraction and suppress transverse deformation. This spatially heterogeneous strain distribution decouples the longitudinal and transverse mechanical responses of the composite, yielding a near-zero effective Poisson's ratio. Critically, the use of line patterns, rather than complex geometric lattices or metamaterial unit cells, provides a mechanically robust and predictable suppression mechanism that does not rely on intricate structural deformation modes and is therefore less sensitive to fabrication imperfections or geometric tolerances. The fabrication process itself is markedly simpler than approaches requiring multi-step lithographic patterning of dissimilar rigid and soft materials: the entire substrate is realized through sequential molding and curing of two PDMS formulations, accessible with standard soft-lithography infrastructure and requiring no specialized bonding treatments. This all-PDMS heterogeneous modulus strategy offers a compelling combination of mechanical performance, optical transparency, processing simplicity, and scalability, establishing a practical and robust platform for next-generation stretchable display technologies.

## 2. Results

Figure 1 illustrates the design rationale and deformation mechanism of the proposed all-PDMS heterogeneous-modulus substrate engineered to achieve near-zero effective Poisson's ratio, thereby enabling distortion-free stretchable display operation. A soft, transparent, and mechanically compliant display platform capable of undergoing large tensile deformation while preserving the spatial fidelity of the displayed content is desirable in wearable electronics (Figure 1a). Unlike conventional elastomeric substrates, which exhibit pronounced lateral contraction under uniaxial strain, the proposed system is designed to maintain geometric stability during stretching. Figure 1b depicts the structural configuration of the heterogeneous substrate, consisting of periodically embedded high-modulus PDMS line patterns within a continuous low-modulus PDMS matrix. The key geometric parameters include the pattern period (P=1000 μm), hard-PDMS width (W= 700 μm), hard-PDMS thickness (Th= 1000 μm), and soft-PDMS thickness (Ts=200 μm). This spatial modulation of elastic modulus within a chemically homogeneous system enables controlled mechanical response without introducing interfacial incompatibilities. Figure 1 C, and d compare the deformation behavior of homogeneous and heterogeneous substrates under tensile loading. In the homogeneous soft PDMS substrate (Figure 1c) uniaxial stretching induces significant transverse contraction due to the intrinsically high Poisson's ratio (~0.5) of elastomers, resulting in lateral shrinkage, distortion of the displayed pattern, and non-uniform strain distribution. This leads to visible deformation of the "ECE35" pattern, representative of pixel distortion in practical displays. In contrast, the heterogeneous substrate (Figure 1 d) exhibits markedly different mechanical behavior. The embedded hard PDMS lines act as mechanically constraining elements that suppress lateral contraction perpendicular to the applied strain. Concurrently, the surrounding soft PDMS regions function as strain absorbers, localizing most of the deformation and accommodating elongation. This cooperative interaction generates a spatially heterogeneous strain field in which longitudinal extension is decoupled from transverse contraction, effectively reducing the overall Poisson's ratio toward zero. As a result, the displayed pattern retains its original aspect ratio with minimal distortion even under significant tensile strain. Additionally, Figure 1d highlights the role of the hard PDMS lines as "assisting stress absorbers," which distribute mechanical load and mitigate stress concentration, thereby enhancing mechanical stability and durability under repeated deformation.

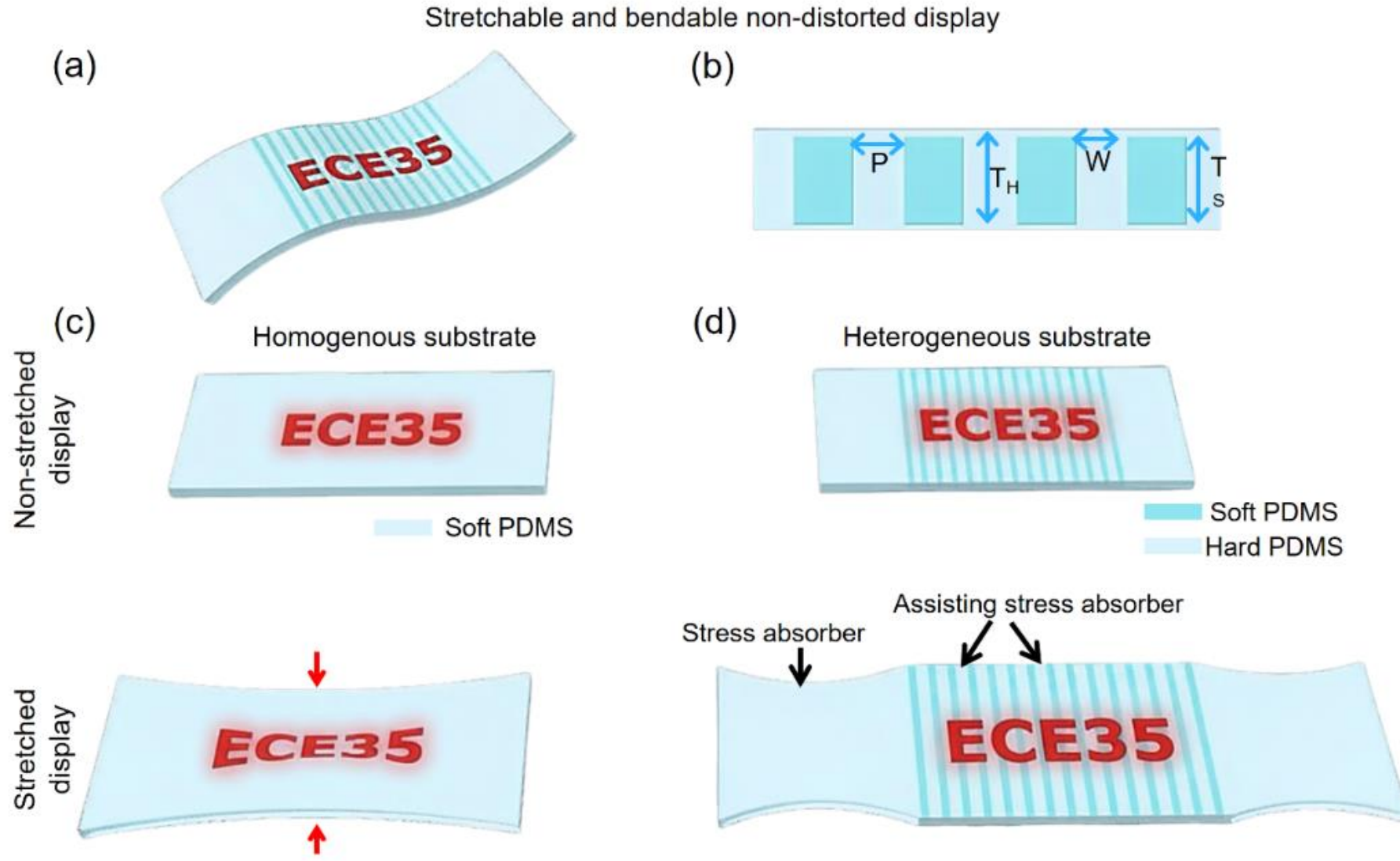

Fig.1: Conceptual illustration of a highly stretchable substrate for distortion-free display. (a) schematic of the proposed all-PDMS-based soft, bendable, and stretchable display, (b) side view of the heterogeneous substrate where pattern dimensions are shown. ‘P’, ‘TH’, W’, and ‘TS’ represent period, hard-PDMS thickness, width, and soft-PDMS thickness, respectively.

Figure 2 presents the finite element modeling (FEM) analysis used to evaluate the influence of heterogeneous modulus patterning on the stress distribution within the proposed substrate. The results provide quantitative insight into how geometric modulation of hard/soft PDMS domains governs local stress localization and redistribution under applied tensile loading. Figure 2a illustrates the computational model and the cross-sectional analysis plane used for extracting stress profiles. The heterogeneous substrate consists of periodically arranged high-modulus PDMS regions embedded within a low-modulus PDMS matrix, consistent with the structural design introduced in Fig. 1. The analysis plane is oriented perpendicular to the patterned lines, enabling evaluation of stress variation across alternating hard and soft domains. The simulated von Mises stress distributions and corresponding line profiles along the longitudinal direction for different heterogeneous pattern ratios, denoted as H1:S1 through H1:S4 (Figure 2 b (i-v)). These ratios represent the relative width of the soft (low-modulus) PDMS regions with respect to the hard (high-modulus) PDMS lines within one periodic unit. In the homogeneous PDMS case (Figure 2 b(i)), the stress distribution is relatively uniform across the substrate, with minor edge concentration effects arising from boundary conditions. This uniformity reflects the absence of mechanical heterogeneity and results in a broadly distributed strain field, consistent with conventional elastomeric behavior. In contrast, the heterogeneous configurations (Figure 2b (ii–v)) exhibit pronounced spatial modulation of stress, strongly correlated with the modulus contrast and geometric ratio. For the H1:S1 case (Figure 2b (ii)), where the widths of hard and soft regions are same, significant stress localization occurs at the interfaces between hard and soft PDMS domains. The stress peaks at the edges of the stiff regions indicate mechanical constraint imposed by the high-modulus segments, while the soft regions experience oscillatory stress variations associated with strain accommodation. As the relative width of the soft PDMS increases (H1:S2 to H1:S4), a progressive redistribution of stress is observed. Specifically, the amplitude of stress oscillations within the patterned region decreases, and the stress profile becomes more homogenized within the soft domains. This indicates that wider soft regions act as more effective strain absorbers, reducing stress concentration at the hard–soft interfaces and facilitating smoother stress transfer across the structure. Simultaneously, the hard PDMS regions continue to serve as mechanical constraints that suppress excessive deformation, maintaining structural integrity.

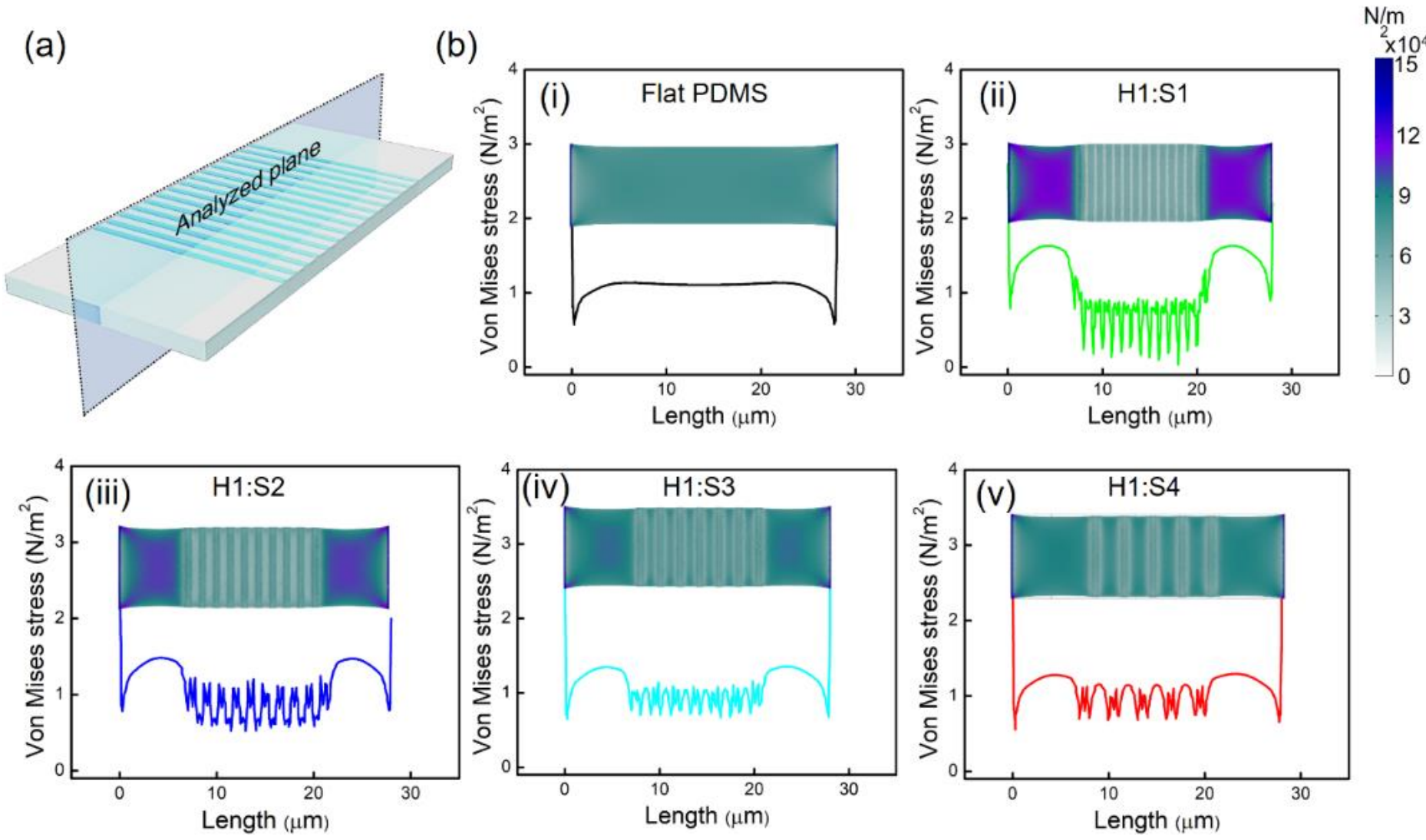

Fig.2: Optimizing the stress distribution based on heterogeneous modulus pattern ratio. (a) showing the cross-sectional cut plane for the stress analysis of the heterogeneous substrate, (b) FEM simulation results demonstrating local von-Mises stress profiles based on the heterogeneous modulus substrates. Ratios correspond to the width of low modulus interspacing with respect to high modulus channels in the patterned region.

Figure 3 illustrates the evolution of transverse (Y-direction) displacement in response to uniaxial stretching, highlighting the role of modulus heterogeneity in suppressing lateral contraction. The cross-sectional analysis plane is defined as perpendicular to the patterned hard PDMS lines at the edge of the device where the deformation occurs the most, enabling direct evaluation of displacement variation across alternating hard and soft regions. The homogeneous PDMS substrate exhibits a smooth and continuous displacement profile with large lateral contraction across the entire width, consistent with the intrinsic near-incompressible behavior of elastomers. In contrast, the introduction of periodically patterned high-modulus PDMS regions leads to pronounced spatial modulation of displacement. For the H1:S1 configuration, lateral displacement is significantly suppressed in the central region, although sharp gradients appear at the interfaces due to stiffness mismatch. With increasing soft-to-hard width ratio (H1:S2 to H1:S4), the displacement profile becomes progressively more uniform, and oscillatory variations within the patterned region are reduced. The soft PDMS domains accommodate most of the deformation, while the hard PDMS lines act as mechanical constraints that limit transverse contraction. As a result, the central region maintains a nearly constant width with minimal lateral shrinkage. These results demonstrate that geometric tuning of the heterogeneous modulus pattern enables effective suppression of transverse displacement, thereby reducing lateral strain and approaching near-zero effective Poisson's ratio behavior, which is critical for maintaining dimensional stability in stretchable display systems. H1:S1 is selected in this work, because it provides the highest density of hard–soft interfaces, maximizing constraint and thus the strongest suppression of lateral deformation. Although this configuration introduces higher stress concentration, these effects are not dominant within the applied strain range (≤30%), where the system remains mechanically stable. Therefore, H1:S1 represents a constraint-dominated reference that enables systematic comparison with higher S ratios to identify an optimal balance between Poisson suppression and strain uniformity.

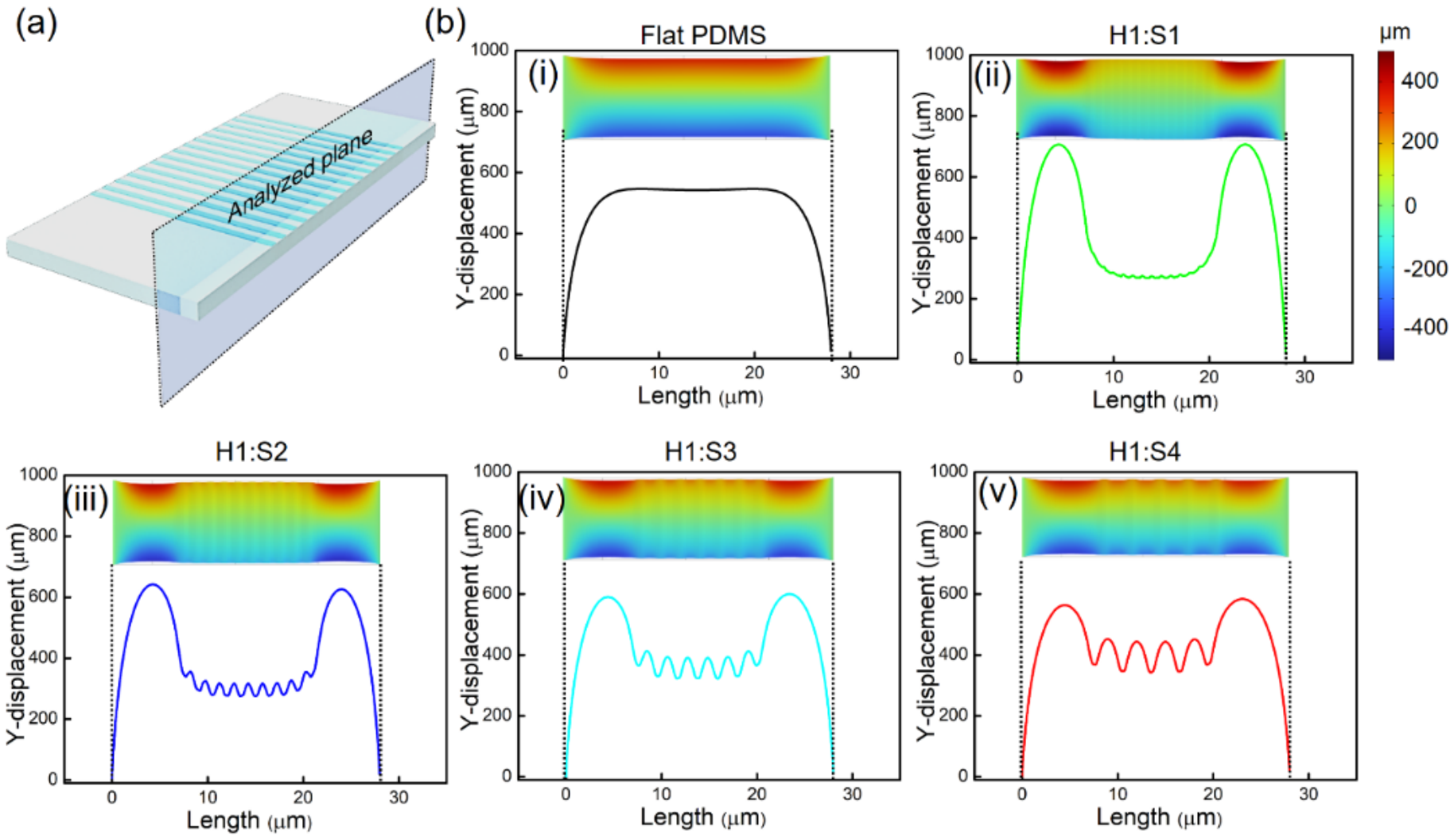

Fig.3: Optimizing the lateral direction displacement based on heterogeneous modulus pattern ratio (a) showing the cross-sectional cut plane for the lateral strain analysis of the heterogeneous substrate, (b) FEM simulation results demonstrating local strain profiles based on heterogeneous modulus substrates. Ratios correspond to the width of low modulus interspacing with respect to high modulus channels in the patterned region.

Figure 4 presents the mechanical response of the heterogeneous-modulus substrate in comparison with homogeneous soft and hard PDMS systems, highlighting its ability to combine mechanical compliance with controlled stress distribution. The simulated axial stress maps show the evolution of stress under increasing tensile strain from 0% to 30%. At low strain, the stress distribution is relatively uniform across the substrate, whereas with increasing strain, stress progressively localizes within the high-modulus PDMS regions. The patterned hard domains sustain higher stress levels, while the surrounding soft PDMS accommodates deformation, resulting in a spatially heterogeneous stress field that mitigates excessive stress accumulation in the central region. The stress–strain curves reveal distinct mechanical behaviors for the three devices (Figure 4b). The homogeneous hard PDMS exhibits a steep, nonlinear increase in stress with strain, indicative of high stiffness and limited deformability. In contrast, the homogeneous soft PDMS shows minimal stress response over the same strain range, reflecting its high compliance. The heterogeneous substrate demonstrates an intermediate response, maintaining low stress levels comparable to soft PDMS at small strains while gradually increasing at higher strains due to load sharing with the embedded hard domains. This behavior indicates effective mechanical coupling between the two phases, enabling both stretchability and structural reinforcement. The extracted Young's modulus further highlights this balance. The homogeneous hard PDMS exhibits the highest modulus, while the soft PDMS shows the lowest (Figure 4c). The heterogeneous substrate possesses an intermediate modulus, confirming that the overall stiffness can be tuned through geometric patterning rather than material substitution. Collectively, these results demonstrate that the heterogeneous-modulus architecture enables decoupling of stiffness and stretchability, providing a mechanically robust yet compliant platform capable of sustaining large deformation while maintaining controlled stress distribution, which is essential for reliable stretchable display operation.

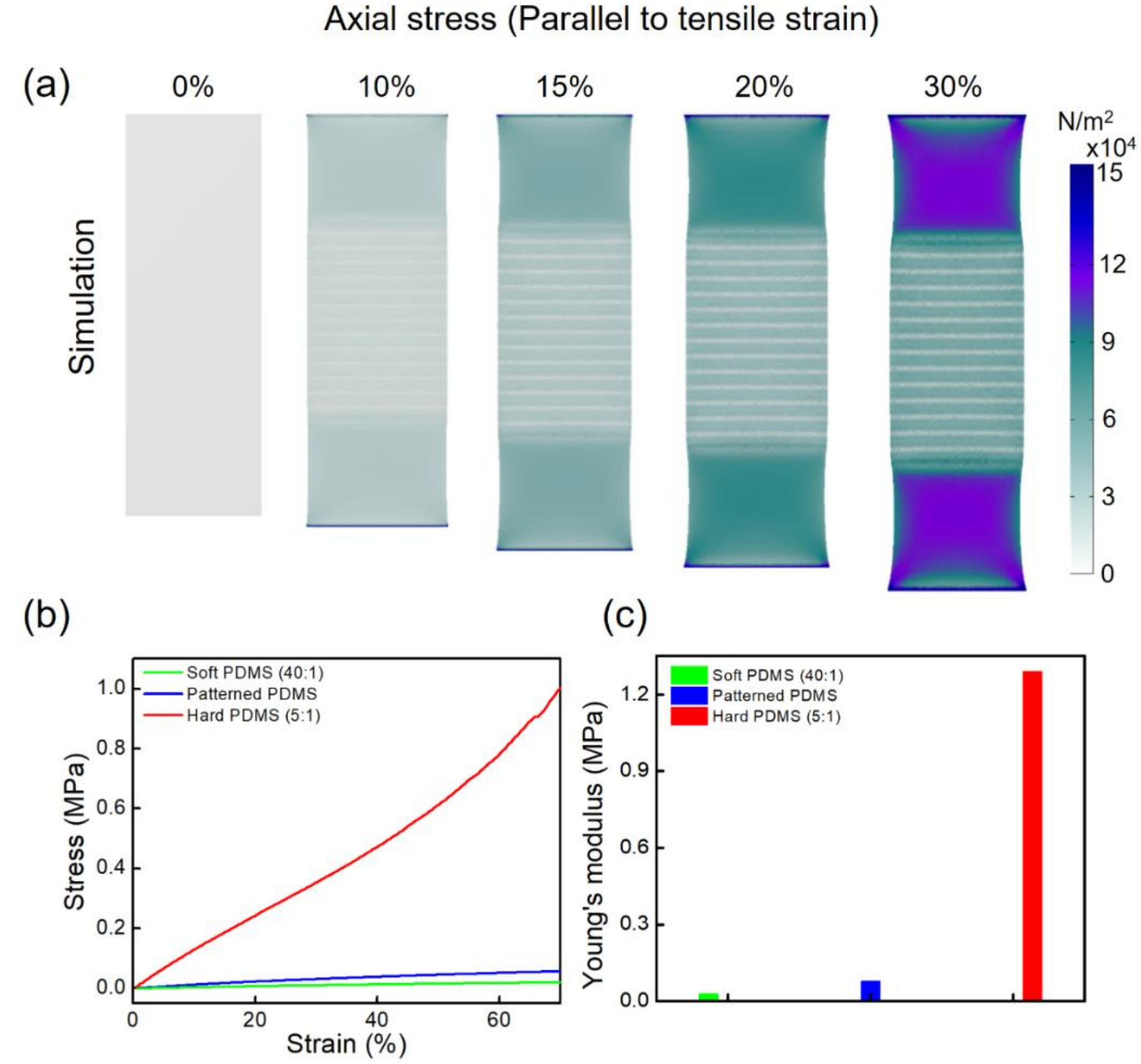

Fig.4: Mechanical characterization of heterogeneous modulus substrates compared with non-patterned substrate (a) FEM results illustrating the Von-Mises stress distribution across a tensile range of 0% to 30%. (b) Comparative stress-strain curves for the homogeneous low-modulus substrate, the homogeneous high-modulus substrate, and the patterned heterogeneous substrate. (c) calculated Young's modulus for the homogeneous low-modulus substrate, the homogeneous high-modulus substrate, and the patterned heterogeneous substrate.

Figure 5 correlates finite element simulations with experimental observations to validate the suppression of transverse deformation in the heterogeneous-modulus substrate under uniaxial tensile loading. The simulated lateral strain maps show the evolution of transverse deformation from 0% to 30% applied strain. At low strain, the displacement field is relatively uniform, whereas increasing tensile strain leads to pronounced spatial gradients, with localized regions of positive and negative lateral displacement (Figure 5b). Notably, the patterned architecture confines transverse deformation, with reduced displacement observed in the central region compared to the edges, indicating effective suppression of lateral contraction. The experimental images exhibit consistent deformation behavior with the simulations (Figure 5b). As tensile strain increases, the substrate elongates while maintaining a nearly constant width in the central region, with only minor lateral contraction near the boundaries. The periodic line patterns remain well-aligned, and no significant distortion or buckling is observed, confirming that the heterogeneous modulus design effectively stabilizes the lateral dimension during stretching. The agreement between simulation and experiment demonstrates that the spatial modulation of modulus enables controlled redistribution of transverse strain, where the soft PDMS accommodates elongation and the embedded hard PDMS structures constrain lateral deformation. This combined behavior leads to reduced transverse displacement and improved dimensional stability, supporting the realization of near-zero effective Poisson's ratio in practical stretchable substrates.

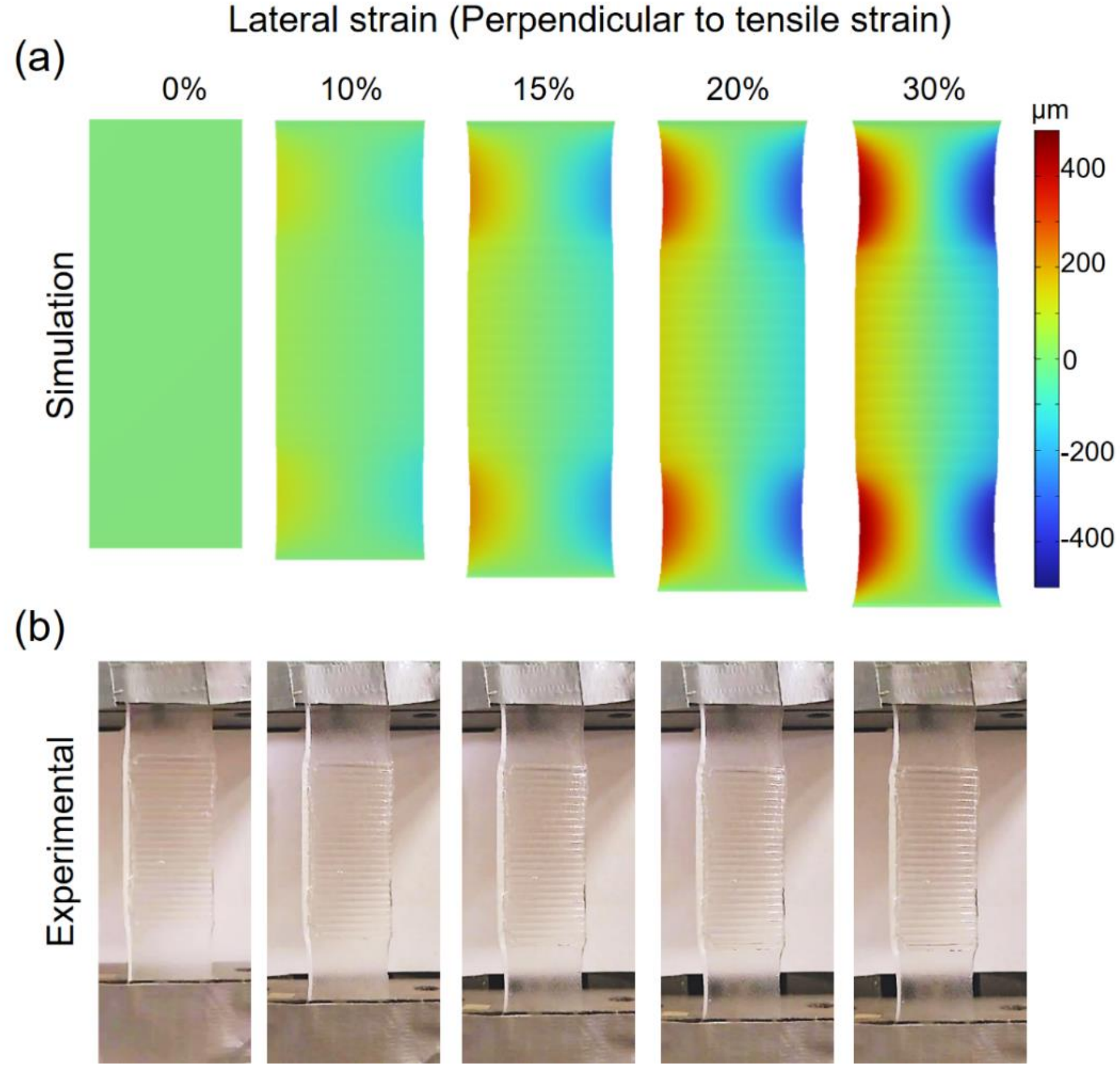

Fig.5: Mechanical characterization of heterogeneous modulus substrates compared with non-patterned substrate a FEM results illustrating the transverse strain distribution for the heterogeneous substrate across a tensile range of 0% to 30%, with the color bar indicating lateral displacement in µm. (b) Corresponding experimental observations of the strain distribution across the heterogeneous substrate under identical tensile loading.

The lateral strain response (Figure 6a) reveals a pronounced difference between the flat and heterogeneous patterned substrates. The homogeneous PDMS exhibits a progressive increase in transverse contraction, with lateral strain reaching approximately −14% at 30% axial strain. In contrast, the patterned PDMS shows significantly suppressed lateral deformation, with lateral strain limited to about −3% at the same axial strain and remaining within ~0 to −3% across the full strain range. The calculated Poisson's ratio (Figure 6b) further quantifies this behavior. The homogeneous PDMS maintains a nearly constant value in the range of ~0.45–0.5, consistent with near-incompressible elastomeric behavior. By comparison, the patterned substrate exhibits a substantially reduced Poisson's ratio of ~0.05–0.1, with minimal variation over the applied strain range (0–30%). These results demonstrate that the heterogeneous modulus architecture effectively reduces transverse contraction by nearly an order of magnitude, enabling near-zero Poisson's ratio behavior and significantly improved dimensional stability under tensile deformation.

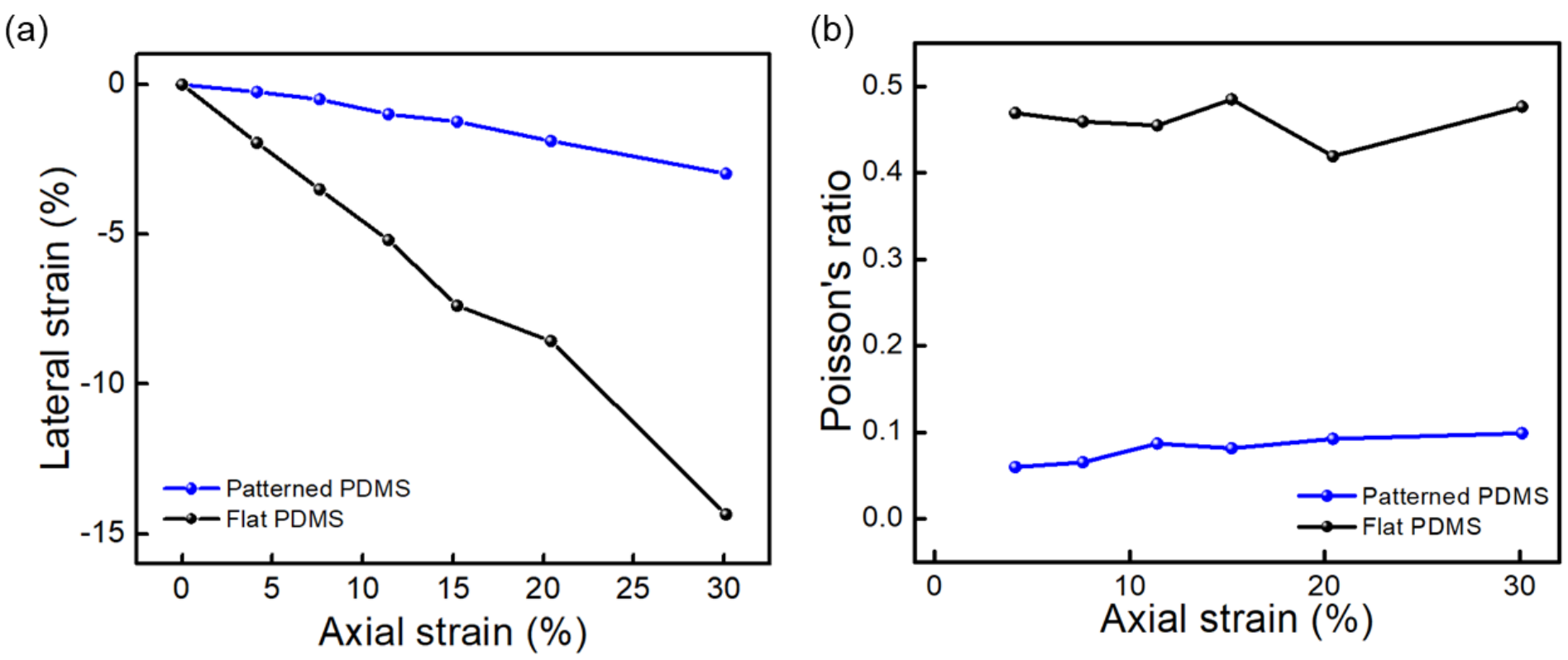

Fig.6: Experimental analysis of lateral strain under different tensile stretching and calculated Poisson's ratio (a) Lateral strain deformation under tensile stress, (b) calculated Poisson's ratio for flat and patterned PDMS.

The optical transmittance spectra (Figure 7) demonstrate that the heterogeneous-modulus substrate maintains high transparency comparable to homogeneous PDMS across the visible range (400–700 nm). The soft PDMS exhibits the highest transmittance, remaining in the range of ~90–92% throughout the spectrum, while the hard PDMS shows slightly lower values of ~88–91%. The patterned PDMS substrate without stretching displays transmittance of ~78–83%, which is marginally reduced compared to homogeneous samples, likely due to light scattering at the modulus-patterned interfaces. Under 30% tensile strain, the patterned substrate shows a slightly higher transmittance (~80–85%), with minimal wavelength dependence. Overall, the variation in transmittance between unstretched and stretched states remains within ~2–3%, indicating that mechanical deformation has negligible impact on optical performance. These results confirm that the heterogeneous modulus design preserves high optical transparency while enabling mechanical stretchability, making it suitable for stretchable display applications.

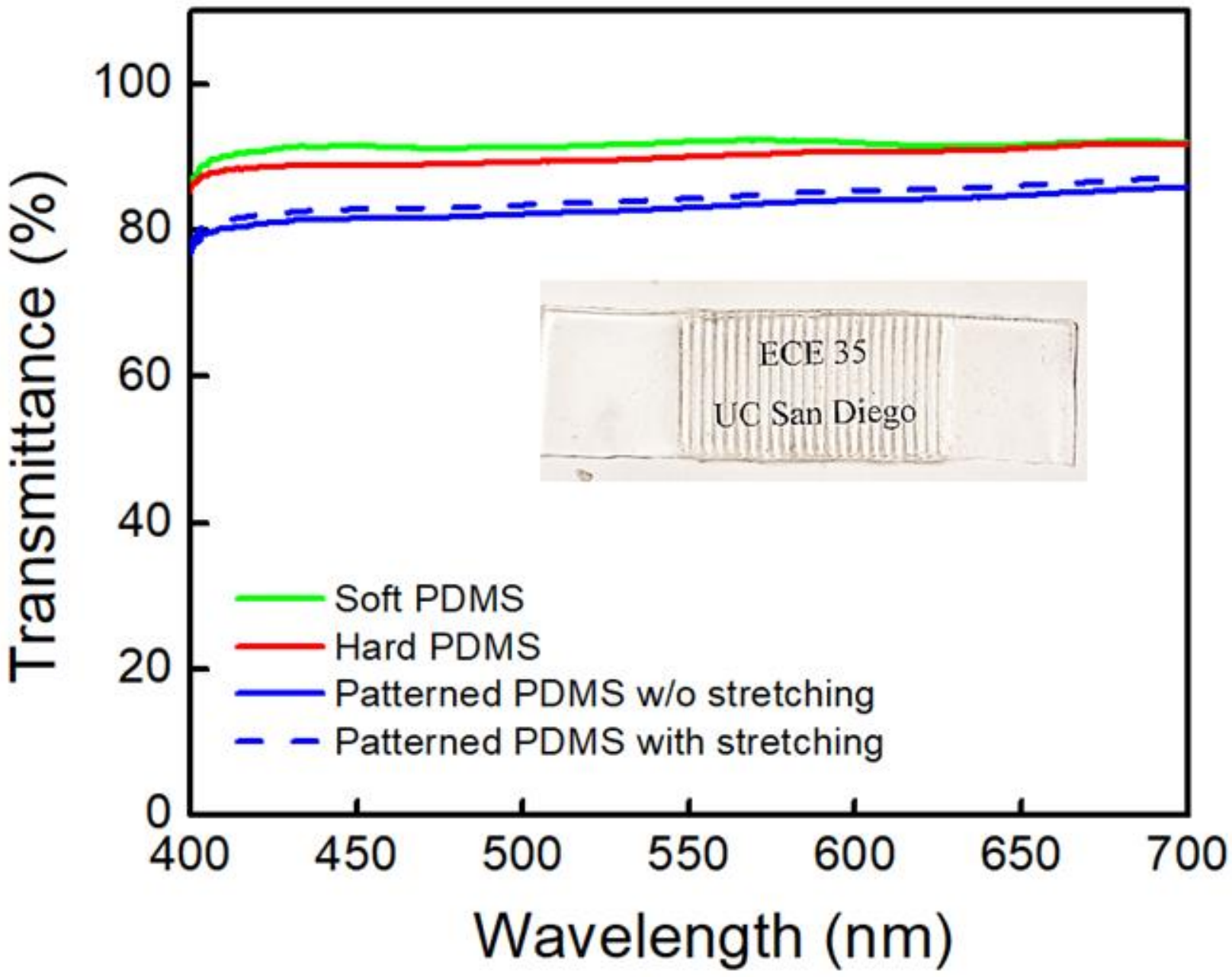

Fig.7: Experimental optical spectra of heterogeneous modulus substrates showing comparison of homogeneous soft PDMS and patterned substrates under 0% and 30% uniaxial tensile strain. Measurements were recorded across the visible spectrum (400–700 nm).

The deformation behavior of LED pixel arrays under uniaxial tensile strain when integrated on homogeneous and heterogeneous PDMS substrates is evaluated (Figure 8). In the homogeneous soft PDMS case (Figure 8a), significant lateral displacement of pixels is observed upon stretching, accompanied pronounced curvature of the device outline. The pixel positions deviate from their initial grid configuration in lateral direction perpendicular to the stretching direction, indicating non-uniformity and distortion consistent with a high Poisson's ratio substrate. In contrast, the patterned heterogeneous substrate exhibits substantially improved geometric stability. The LED pixels maintain nearly uniform spacing along both axial and transverse directions, with minimal lateral displacement even under applied strain. The overall device outline remains comparatively straight, with only slight edge curvature, indicating effective suppression of transverse deformation. This behavior confirms that the heterogeneous modulus architecture effectively preserves pixel alignment and spatial uniformity by decoupling axial strain from lateral contraction. These results directly validate the mechanical design strategy, demonstrating that the near-zero Poisson's ratio behavior of the patterned substrate translates to improved pixel stability and reduced image distortion in functional stretchable display systems.

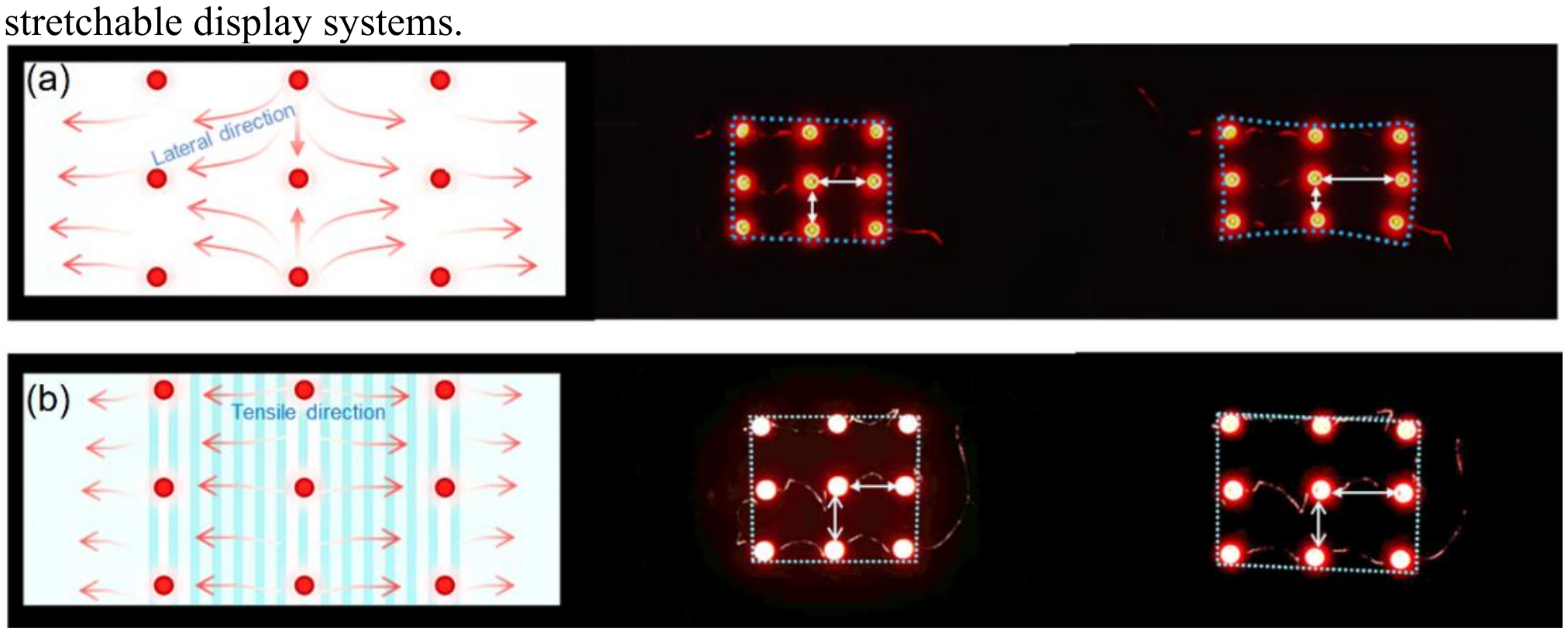

Fig.8: Experimental demonstration of LED array pixel displacements under tensile stretching of substrates. (a) the LED array pixel displacements under tensile stretching integrated onto a uniformly soft PDMS substrate. Under stress, the pixels experienced lateral displacement and warping, indicated by the curvature in the dotted outline. (b) the LEDs on a patterned PDMS substrate exhibited more uniform pixel spacing and minimal curvature in the dotted outline.

## 3. Conclusion

In this work, a heterogeneous-modulus elastomeric substrate composed entirely of PDMS was developed to address the intrinsic limitation of large Poisson's ratio in stretchable displays. By embedding periodically patterned high-modulus PDMS within a soft PDMS matrix, the substrate achieves effective suppression of transverse deformation while preserving mechanical compliance and optical transparency. Both simulation and experimental results consistently demonstrate reduced lateral strain (~−3% at 30% axial strain) and a near-zero Poisson's ratio (~0.05–0.1), in contrast to conventional PDMS (~−14% lateral strain, $\nu \approx 0.45$–$0.5$). The structured modulus distribution enables controlled stress and strain partitioning, minimizing distortion and maintaining dimensional stability under tensile loading. This is further validated through LED array demonstrations, where the patterned substrate preserves pixel alignment and spacing with negligible warping. Importantly, the all-PDMS approach avoids interfacial incompatibility and maintains high optical transmittance (>~80%) even under strain. Overall, this work establishes a simple, scalable strategy for achieving near-zero Poisson's ratio behavior in elastomeric substrates, providing a robust platform for distortion-free stretchable optoelectronic systems.

## 4. Experimental Methods

### *4.1 Fabrication of Zero-Poisson's ratio substrate*

The heterogeneous-modulus substrate was fabricated through a sequential soft-lithography and molding process using two PDMS formulations with different crosslinking ratios. First, a microstructured mold containing periodic line patterns was prepared using a Bambu A1 Mini (Bambu Lab) 3D printer with polylactic acid silk (PLA silk) filament. A high-modulus PDMS prepolymer (e.g., 5:1 base-to-curing-agent ratio) was cast onto the mold and cured to form patterned hard PDMS lines. Excess material was removed to ensure well-defined, isolated line structures. Subsequently, a low-modulus PDMS (e.g., 40:1 ratio) was prepared and poured over the patterned hard PDMS layer to fill the interspacing regions and form a continuous matrix. The bilayer structure was then thermally cured to achieve covalent bonding between the hard and soft PDMS phases, ensuring a mechanically robust and monolithic interface without delamination. After curing, the composite film was carefully peeled from the mold, yielding a flat heterogeneous substrate with embedded hard PDMS line patterns. For device integration, an additional thin layer of soft PDMS was cast on top to planarize the surface and encapsulate the patterned features, resulting in a smooth, transparent, and stretchable platform. This fabrication approach enables precise control over pattern geometry (period, width, and spacing) while maintaining chemical homogeneity, scalability, and compatibility with standard soft-lithography processes.

### *4.2 Fabrication of 3x3 LED Display*

A flexible 3x3 array of red surface-mount micro-LEDs (ASMM-CR03-AS402, Broadcom, peak wavelength 630 nm) was assembled onto the fully cured PDMS substrate to create a stretchable display. The LEDs were connected in rows with 38 AWG enamel-coated copper wires (Remington). Each wire was cut to a pre-extended length that enabled the PDMS substrate to stretch up to 130% of its relaxed length. The ends of the wires were stripped and soldered to the LED terminals using a temperature-controlled soldering iron set to 305oC. Each wire was formed into an arc shape such that the adjacent LEDs were spaced 9.4 mm apart while the rows were spaced 4.5 mm apart. Small drops of uncured PDMS (40:1 ratio) were then applied between the PDMS substrate and the LED array to act as an adhesive and then cured at 80˚ C for 12 h.

### *4.3 Characterization*

To evaluate the optical performance of the homogeneous PDMS and heterogeneous MM substrates, visible light transmittance was characterized using a UV–vis spectrometer (Ocean Optics, USB2000+) integrated with a custom stretching jig. These measurements were conducted across a wavelength range of 400 to 700 nm under both static (0% strain) and tensile (30% strain) conditions. The tensile stress tests were performed on a TA Instruments Discovery HR-30 hybrid rheometer at a constant extension rate of 100 μm $s^{-1}$.

## 5. Back matter

### *5.1 Acknowledgment*

The authors would like to acknowledge financial support from 2024 Alfred P. Sloan Research Fellow, 2023 Beckman Young Investigator Award, from the Arnold and Mabel Beckman Foundation, the Moore Foundation to the PAIR-UP Imaging Science Program, Air Force Office of Scientific Research MURI (Award No. FA9550-22-1-0312), Silicon Valley Community

Foundation (Grant No. DAF2023-331948); cZi Dynamic imaging via the Chan Zuckerberg Donor Advised Fund (DAF) through the Silicon Valley Community Foundation. The work was performed in part at the San Diego Nanotechnology Infrastructure. We thank Photometrics, Inc. (Chico, CA; dissolved 2022) for support made possible through the donation honoring the research legacy of co-founder Brian Pierce (1957–2018).

### *5.2 Disclosures*

These authors contributed equally: Joseph Nguyen, Kim Tran, Tristan Tjussardi.

### *5.3 Data availability statement*

The data that supports the findings of this study are available from the corresponding author upon reasonable request.

### *5.4 Conflict of interest*

The authors declare no competing interests.

## 6. References

### References

1. He J, Wei R, Ma X, et al. Contactless user-interactive sensing display for human-human and human-machine interactions. Adv Mater 2024;36:e2401931.
2. Y.Zhao, F.Lan, Y.Li, et al. "Coaxial Electroluminochromic Fibers with Dynamic RGB Switching for Pixelated Smart Textiles." Adv. Mater.37, no. 37 (2025): 37, 2505012. https://doi.org/10.1002/adma.202505012
3. Choi, H.W., Shin, DW., Yang, J. et al. Smart textile lighting/display system with multifunctional fibre devices for large scale smart home and IoT applications. Nat Commun 13, 814 (2022). https://doi.org/10.1038/s41467-022-28459-6
4. Tomoyuki Yokota et al. Ultraflexible organic photonic skin. Sci. adv .2, e1501856(2016). DOI:10.1126/sciadv.1501856
5. Park, J., Heo, S., Park, K. et al. Research on flexible display at Ulsan National Institute of Science and Technology. npj Flex Electron 1, 9 (2017). https://doi.org/10.1038/s41528-017-0006-9
6. Jeon, C. H., Park, J. W., Kang, B. W., Jang, S. H., Kwon, K. J., Hong, S. K., … Jang, J. (2023). Analysis of heat diffusion considering driving images on 6-inch flexible AMOLED display. Journal of Information Display, 24(3), 169–175. https://doi.org/10.1080/15980316.2023.2169378
7. J. H.Koo, D. C.Kim, H. J.Shim, T.-H.Kim, D.-H.Kim, Flexible and Stretchable Smart Display: Materials, Fabrication, Device Design, and System Integration, Adv. Funct. Mater.2018, 28, 1801834. https://doi.org/10.1002/adfm.201801834
8. Kim, D.W., Kim, S.W., Lee, G. et al. Fabrication of practical deformable displays: advances and challenges. Light Sci Appl 12, 61 (2023). https://doi.org/10.1038/s41377-023-01089-3
9. Lee, Y., Guan, W., Hsieh, E.Y. and Nam, S. (2024), Opportunities for Nanomaterials in Stretchable and Free-Form Displays. Small Sci., 4: 2300143. https://doi.org/10.1002/smsc.202300143
10. Myung Sub Lim, Minwoo Nam, Seungyeop Choi, Yongmin Jeon, Young Hyun Son, Sung-Min Lee, and Kyung Cheol Choi, Nano Letters 2020 20 (3), 1526-1535, DOI: 10.1021/acs.nanolett.9b03657
11. Yeongjun Lee et al. Standalone real-time health monitoring patch based on a stretchable organic optoelectronic system. Sci. Adv. 7, eabg9180 (2021). DOI:10.1126/sciadv.abg9180
12. J.Byun, S.Chung, Y.Hong, Artificial Soft Elastic Media with Periodic Hard Inclusions for Tailoring Strain-Sensitive Thin-Film Responses, Adv. Mater.2018, 30, 1802190. https://doi.org/10.1002/adma.201802190
13. D.Yin, N.-R.Jiang, Z.-Y.Chen, Y.-F.Liu, Y.-G.Bi, X.-L.Zhang, J.Feng, H.-B.Sun, Roller-Assisted Adhesion Imprinting for High-Throughput Manufacturing of Wearable and Stretchable Organic Light-Emitting Devices. Adv. Optical Mater.2020, 8, 1901525. https://doi.org/10.1002/adom.201901525
14. Chen Du, Yiqiang Wang, and Zhan Kang, Auxetic Kirigami Metamaterials upon Large Stretching, ACS Applied Materials & Interfaces 2023 15 (15), 19190-19198, DOI: 10.1021/acsami.3c00946
15. Greaves, G., Greer, A., Lakes, R. et al. Poisson's ratio and modern materials. Nature Mater 10, 823–837 (2011). https://doi.org/10.1038/nmat3134

16. Roderic Lakes, Foam Structures with a Negative Poisson's Ratio. Science 235, 1038-1040(1987). DOI:10.1126/science.235.4792.1038
17. Young-Joo Lee, Seung-Min Lim, Seol-Min Yi, Jeong-Ho Lee, Sung-gyu Kang, Gwang-Mook Choi, Heung Nam Han, Jeong-Yun Sun, In-Suk Choi, Young-Chang Joo, Auxetic elastomers: Mechanically programmable meta-elastomers with an unusual Poisson’s ratio overcome the gauge limit of a capacitive type strain sensor,Extreme Mechanics Letters, , 31 (2019), p. 100516.
18. Abdoulaye Ndao, Edwin B. Fohtung, Moussa N'Gom, Thomas A. Searles, Kimani Toussaint, Yanne K. Chembo; Synergistic integration of metasurfaces and quantum photonics: Pathways to next-generation technologies. Appl. Phys. Rev. 1 December 2025; 12 (4): 041318. https://doi.org/10.1063/5.0226259
19. Abdoulaye Ndao, Roland Salut, Miguel Suarez, Fadi I Baida, Plasmonless polarization-selective metasurfaces in the visible range, J. Opt. 20 045003, 2018
20. Taylor, M., Francesconi, L., Gerendás, M., Shanian, A., Carson, C. and Bertoldi, K. (2014), Low Porosity Metallic Periodic Structures with Negative Poisson's Ratio. Adv. Mater., 26: 2365-2370. https://doi.org/10.1002/adma.201304464
21. R. Libanori, R. M. Erb, A. Reiser, H. Le Ferrand, M. J. Suess, R. Spolenak, A. R. Studart, Nat. Commun. 2012, 3, 1265.
22. Jun-Hee Park, Ashok Kodigala, Abdoulaye Ndao, and Boubacar Kanté, "Hybridized metamaterial platform for nano-scale sensing," Opt. Express 25, 15590-15598 (2017)
23. Smith, C. W., Grima, J. N. & Evans, K. E., A novel mechanism for generating auxetic behaviour in reticulated foams: missing rib foam model. Acta. Mater. 48, 4349–4356 (2000).
24. Fukumoto, A. First-principles pseudopotential calculations of the elastic properties of diamond, Si, and Ge. Phys. Rev. B 42, 7462–7469 (1990).
25. Byun J, Chung S, Hong Y. Artificial soft elastic media with periodic hard inclusions for tailoring strain-sensitive thin-film responses. Adv Mater 2018;30:e1802190.
26. A Ndao, J Salvi, R Salut, M-P Bernal, T Alaridhee, A Belkhir and F I Baida, Resonant optical transmission through sub-wavelength annular apertures caused by a plasmonic transverse electromagnetic (TEM) mode, 2014 J. Opt. 16 125009.
27. Saeed Hemayat, Liyi Hsu, Jeongho Ha, and Abdoulaye Ndao, "Near-unity uniformity and efficiency broadband meta-beam-splitter/combiner," Opt. Express 31, 3984-3997 (2023).
28. Yin L, Lv J, Wang J. Structural innovations in printed, flexible, and stretchable electronics. Adv Mater Technol 2020;5:2000694.
29. Jeongho Ha, Abdoulaye Ndao, Liyi Hsu, Jun-Hee Park, and Boubacar Kante, "Planar dielectric cylindrical lens at 800 nm and the role of fabrication imperfections," Opt. Express 26, 23178-23184 (2018)
30. Jun-Hee Park, Jeongho Ha, Liyi Hsu, et al. "Observation of robust subwavelength phase singularity in chiral medium," *Advanced Photonics* 7(3), 035001 (17 Mar 2025) https://doi.org/10.1117/1.AP.7.3.035001
31. A Ndao, J Salvi, R Salut, MP Bernal, T Alaridhee, A Belkhir, FI Baida, Journal of Optics 16 (12), 125009
32. A Nauman, JC Choi, YM Cho, JW Lee, JH Na, HR Kim, Materials & Design, 237, 112515, 2024
33. Lee Y, Lim S, Yi S, et al. Auxetic elastomers: mechanically programmable meta-elastomers with an unusual Poisson’s ratio overcome the gauge limit of a capacitive type strain sensor. Extreme Mech Lett 2019;31:100516.
34. Nauman A, Khaliq HS, Choi JC, Lee JW, Kim HR. Topologically engineered strain redistribution in elastomeric substrates for dually tunable anisotropic plasmomechanical responses. ACS Appl Mater Interfaces 2024;16:6337-47.
35. J.-C.Choi, H. Y.Jeong, J.-H.Sun, J.Byun, J.Oh, S. J.Hwang, P.Lee, D. W.Lee, J. G.Son, S.Lee, S.Chung, Bidirectional Zero Poisson's Ratio Elastomers with Self-Deformable Soft Mechanical Metamaterials for Stretchable Displays. Adv. Funct. Mater.2024, 34, 2406725. https://doi.org/10.1002/adfm.202406725
36. Choi, J.-C., Lee, D.W., Jeong, H.Y., Yang, J., Jeong, S., Kim, W., Son, J.G., Kim, H., Hong, Y. and Chung, S. (2025), Depth-Modulus Engineered Meta-Elastomers for Multiaxial Strain Programming in Stretchable Displays. Small Struct., 6: 2400513. https://doi.org/10.1002/sstr.202400513.
37. Choi, J.-C., Lee, D.W., Jeong, H.Y., Yang, J., Jeong, S., Kim, W., Son, J.G., Kim, H., Hong, Y. and Chung, S. (2025), Depth-Modulus Engineered Meta-Elastomers for Multiaxial Strain Programming in Stretchable Displays. Small Struct., 6: 2400513. https://doi.org/10.1002/sstr.202400513
38. Park, D. H., Lee, D. W., Jeong, H. Y., Choi, J. C., & Chung, S. (2025). Auxetic mechanical metamaterials: efficient strain engineering for highly reliable free-form displays. Journal of Information Display, 26(4), 447–460. https://doi.org/10.1080/15980316.2025.2539686
39. Gao, Y.; Yu, L.; Yeo, J. C.; Lim, C. T. Flexible Hybrid Sensors for Health Monitoring: Materials and Mechanisms to Render Wearability. Adv. Mater. 2020, 32, 1902133 DOI: 10.1002/adma.201902133
40. K. Bertoldi, V. Vitelli, J. Christensen, M. V. Hecke, Nat. Rev. Mater. 2017, 2, 1.
41. Asad Nauman, Jun-Chan Choi, Jae-Won Lee, Young-Min Cho, Min-Seok Kim, Bilal ud din Khan, Dong Yeol Hyeon, HakSu Jang, Kwi-Il Park, Hak-Rin Kim, Stretchable anisotropic piezoelectric sensors with modulus engineering for monitoring multiaxial joint motion, Materials & Design, 2025, 254 , 114120 https://doi.org/10.1016/j.matdes.2025.114120.